# Many-body resonance in a correlated topological kagome antiferromagnet

Songtian S. Zhang[1*], Jia-Xin Yin[1*†], Muhammad Ikhlas[2], Hung-Ju Tien[3], Rui Wang[4], Nana Shumiya[1], Guoqing Chang[1], Stepan S. Tsirkin[5], Youguo Shi[6,7], Changjiang Yi[6,7], Zurab Guguchia[8], Hang Li[6], Wenhong Wang[6], Tay-Rong Chang[3], Ziqiang Wang[9], Yi-Feng Yang[6,7], Titus Neupert[5], Satoru Nakatsuji[2,10,11], M. Zahid Hasan[1,12†]

**Affiliations:**
[1]Laboratory for Topological Quantum Matter and Advanced Spectroscopy, Department of Physics, Princeton University, Princeton, NJ, USA.
[2]Institute for Solid State Physics, University of Tokyo, Kashiwa 277-8581, Japan.
[3]Department of Physics, National Cheng Kung University, Tainan 701, Taiwan
[4]National Laboratory of Solid State Microstructures and Department of Physics, Nanjing University, Nanjing 210093, China
[5]Department of Physics, University of Zurich, Zurich, Switzerland.
[6]Beijing National Laboratory for Condensed Matter Physics and Institute of Physics, Chinese Academy of Sciences, Beijing 100190, China
[7]University of Chinese Academy of Sciences, Beijing 100049, China
[8]Laboratory for Muon Spin Spectroscopy, Paul Scherrer Institute, Villigen PSI, Switzerland
[9]Department of Physics, Boston College, Chestnut Hill, Massachusetts 02467, USA.
[10]Department of Physics, University of Tokyo, Hongo, Bunkyo-ku, Tokyo 113-0033, Japan,
[11]CREST, Japan Science and Technology Agency (JST), 4-1-8 Honcho Kawaguchi, Saitama 332-0012, Japan
[12]Lawrence Berkeley National Laboratory, Berkeley, California 94720, USA.
*These authors contributed equally.
†Corresponding authors. Email: jiaxiny@princeton.edu, mzhasan@princeton.edu

**We use scanning tunneling microscopy/spectroscopy (STM/S) to elucidate the atomically resolved electronic structure in strongly correlated topological kagome antiferromagnet $Mn_3Sn$. In stark contrast to its broad single-particle electronic structure, we observe a pronounced resonance with a Fano line shape at the Fermi level resembling the many-body Kondo resonance. We find that this resonance does not arise from the step edges or atomic impurities, but the intrinsic kagome lattice. Moreover, the resonance is robust against the perturbation of a vector magnetic field, but broadens substantially with increasing temperature, signaling strongly interacting physics. We show that this resonance can be understood as the result of geometrical frustration and strong correlation based on the kagome lattice Hubbard model. Our results point to the emergent many-body resonance behavior in a topological kagome magnet.**

Studying the effects of correlation and topology in quantum materials is emerging as one of the central themes in condensed matter physics [1]. In electron systems with strong Coulomb interaction, they often exhibit exotic electronic and magnetic properties that cannot be sufficiently accounted for by the non-interacting properties of their individual constituents. As such, the realization of these emergent properties in topological materials can lead to unpredicted manifestations of their many-body physics. Recently, a series of correlated kagome magnets have been observed to have anomalous transport response, correlated topological electronic structure and giant spin-orbit tunability [2-16]. Among these, $Mn_3Sn$ stands out due to its antiferromagnetism and the absence of any non-kagome layers. It is also one of the rare antiferromagnets that exhibits large anomalous Hall and Nernst effects, arising from the Berry curvature due to gapped magnetic nodal lines leading to Weyl fermions [2,3]. In addition to its large Berry curvature, photoemissions experiments have shown a broad single-particle electronic structure with a band renormalization factor as large as five [4]. Therefore, this combination of kagome lattice, topological bands and strong correlation in Mn₃Sn lends itself to be a fascinating platform for studying the strongly correlated topological kagome magnet.

Mn₃Sn crystallizes in space group P6₃/mmc with the lattice constant $a = b = 5.7$ Å, where each layer consists of a kagome lattice made of Mn atoms stuffed with Sn atoms [Fig. 1(a)], with AB stacking of the equivalent layers [Fig. 1(b)]. A first-principles calculation of the bulk band structure [Fig. 1(c)] finds several flat bands and Weyl fermions [3,4]. Due to strong interplane coupling, the as-cleaved surface does not exhibit an atomic lattice structure [5,12]. To prepare atomic surfaces, we first cleave the sample and then anneal it at 1100K for one hour. With this *in-situ* annealing process, we can measure large clean flat areas with terraces larger than 300nm in size [Fig. 1(d)]. We directly visualize the Mn₃Sn atomic surface at 4.6K with hexagonal symmetry and expected lattice constant, as shown in Fig. 1(e). A comparison to simulation finds reasonable agreement [Fig. 1(e) inset], similar to the kagome surfaces in Fe₃Sn₂ [5,13] and Co₃Sn₂S₂ [10,15]. An analysis of a line cut taken across several single atomic steps finds each step to be approximately 2.3Å, consistent with its half c-axis unit [Fig. 1(f)]. Further analysis across a single atomic step reveals the interlayer alignment of the kagome lattice, consistent with the AB stacking of the bulk crystal structure [Fig. 1(g)].

Having visualized the atomic kagome surface of Mn₃Sn, we now investigate its low-energy electronic structure. Measuring the dI/dV spectrum along a line of 50nm, far from any step edge, we find a consistent resonance feature at $E_F$ with a line-shape asymmetry [Fig. 2(a)]. Near the atomic step edge, we find that the resonance is strongly suppressed [Fig. 2(b) and its inset], demonstrating that the resonance is not from the breaking bonding induced scattering from the step edge. As this peak is observed at every position on the kagome surface, this resonance state is not related to an isolated impurity but is a feature of the kagome lattice. This point is further supported in Fig. 2(c), that the resonance is slightly suppressed by the isolated atomic impurity. The spectral asymmetry of this resonance leads us to consider the Fano equation $F(E) \approx \frac{(q+E/\Gamma)^2}{1+(\frac{E}{\Gamma})^2}$, where $q$ is the quality factor that quantifies the coupling of the tip to the discrete state and $\Gamma$ is the resonance width. The spatially averaged dI/dV spectrum can indeed be fitted by A*$F$(E)+B as shown in Fig. 2(d), where A and B are the additional adjustment parameters. It can also be seen that there is a non-zero background, indicative that not all states are associated with the resonance. $\Gamma = 3.9$meV is the characteristic energy scale of the resonance. If we associate this state with the Kondo lattice resonance, the estimated Kondo temperature would amount to $T_K = \Gamma/1.4k_B = 32$K.

A magnetic field dependent measurement finds that despite some weak broadening the resonance peak does not split or shift under an out-of-plane 4T field or in-plane 2T field relative to the unperturbed resonance state [Fig. 3(a) orange, brown, and blue curves respectively]. In reference to tunneling experiments in YbRh₂Si₂ and SmB₆ among others [6,7,17] where a Kondo resonance with a Fano line shape was observed, the resonance was also not split by a strong magnetic field. In contrast to the weak field response, we observe a strong temperature dependence of the resonance. Our temperature-dependent measurement finds that the resonance peak is substantially suppressed and broadens with increasing temperature [Fig. 3(b), solid lines]. In tunneling experiments, the dI/dV spectra measure the convolution of the DOS and the derivative of the Fermi Dirac distribution function. As such, we also plot the temperature convolution of the spectrum taken at our lowest temperature (T=4.6K) for each temperature (dotted lines) for comparison. We see that the actual data shows a stronger temperature broadening effect than the convoluted curves, indicating an interaction driven resonance with intrinsic temperature dependence. Next, we fit the experimental data with the thermally convoluted Fano function in Fig. 3(c). During the fitting, we find that although $q$ shows little variation, $\Gamma$ increases substantially with T, which is associated with the intrinsic thermal broadening of the DOS peak. The spatially independent line-shape, magnetic field response, and thermal broadening of this resonance all resemble the behavior of the resonance in systems that can be described by the Kondo lattice model [6,7,17-20]. Fitting $\Gamma(T)$ with the extended phenomenological expression derived initially for the single-impurity model [21,22,23], $\Gamma = (2(k_B T_K)^2 + (\pi k_B T)^2)^{0.5}$, we obtain an estimated Kondo temperature of $T_K = 30$K as shown in Fig. 3(d). The obtained $T_K$ is consistent with the energy scale of the linewidth at base temperature. In the bulk resistivity data [24], we have also observed a characteristic Kondo upturn around this $T_K$ determined by STM/S, indicating its bulk origin. We can use this estimated Kondo temperature to re-examine the non-splitting nature of the resonant state under the magnetic field. The minimum field with which the resonant state splits is related to the Kondo temperature by approximately $g\mu_B B \approx 0.5 k_B T_K$ [25,26]. For an estimated $T_K = 30K$ and $g \approx 2$, a minimum field of approximately 11 T would be required. For the smaller fields applied in our experiment, some broadening of the resonance can be observed, but without clearly splitting.

Indeed, there are several ways in which the crystal and electronic structure of $Mn_3Sn$ can be considered analogous to systems described by the Kondo model [27]. In the Kondo model, the localized moments in a lattice are screened by the itinerant conductive sea, forming periodic singlet states [Fig. 4(a)]. In their excitation spectrum, the many-body interactions between the flat band and itinerant conduction band can manifest as a resonance with a Fano line shape at the Fermi level [Fig. 4(b)]. This behavior is mostly observed in heavy fermion systems with localized $f$ orbitals. However, the key components for the formation of such a resonance—namely the flat band and the strong Coulomb interactions—can in principle, be satisfied in a strongly correlated $3d$ kagome metal. In this case, the kagome lattice localizes the electronic wavefunctions in place of the heavy $f$ electrons, with the itinerant conduction sea naturally arising from its metallic state. Although the higher energy single-particle flat band cannot be clearly resolved in this strongly correlated system due to the short quasi-particle lifetime [4], the flat band is a general feature of the kagome lattices as demonstrated in its first-principles calculations. Theoretically, we consider a set of kagome flat bands touching a dispersive band with a dominating Hubbard U and an inter-band coupling [Fig. 4(c)]. By solving the coupled equation of motion for the Green's function derived to the third order for the kagome lattice Hubbard model [24], we show that there can be a many-body resonance at the Fermi level in the density of states [Fig. 4(d)].

Moreover, in contrast to heavy fermion materials, $Mn_3Sn$ exhibits anomalous Hall and Nernst effects coming from Berry curvature with topological fermions and Fermi arcs[2,4,11], and the interplay between frustrated magnetism, Berry-phase and many-body effects within this material has the potential to open new research directions. The achievement of a large atomic flat surface in this material by our methodology also indicates the possibility of engineering such stoichiometric materials down to atomically thin layers to realize the quantized anomalous effect towards the realization of high-temperature interacting dissipationless modes. Finally, it has not escaped our attention that previous tunneling data directly into the kagome layer of $Fe_3Sn_2$ and $Co_3Sn_2S_2$ all exhibit an anomalous zero-bias peak [5,10,13,15], therefore this many-body resonance phenomenon may be ubiquitous in this family of strongly correlated kagome metals.

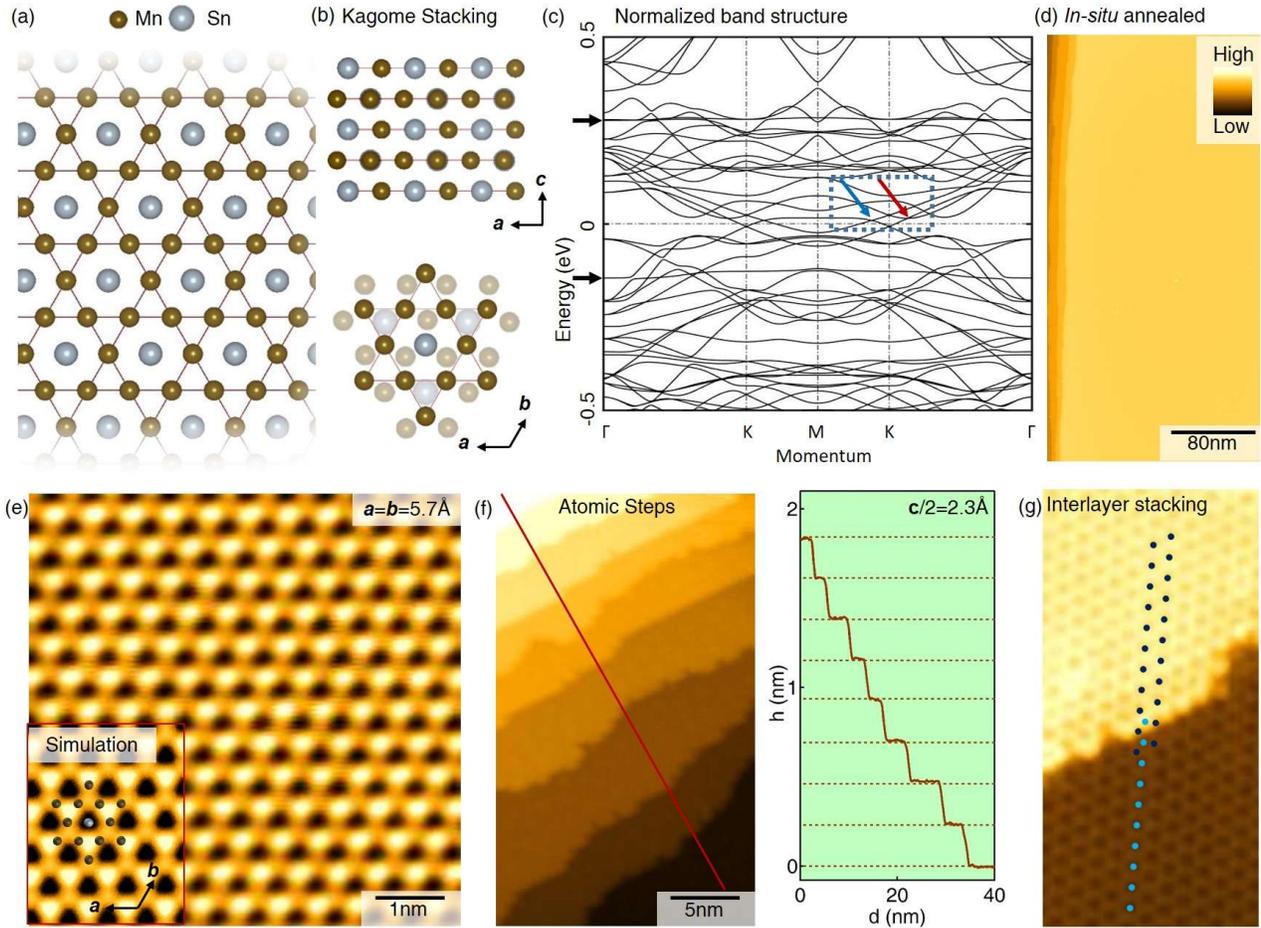

Fig. 1 (a) Crystal structure of the Mn$_3$Sn kagome lattice. (b) c-axis AB stacking of the kagome lattice from side view and top view, respectively. (c) Normalized bulk band structure of Mn$_3$Sn in the conformal Brillouin zone showing low energy flat bands (black arrows) and Weyl nodes with two examples indicated by red and blue arrows. (d)(e) STM topographic images at different scales of Mn$_3$Sn after annealing the cleaved surface at 1100K for 1 hour, showing a large atomic flat surface showing the lattice with hexagonal symmetry. Inset of (e): first-principles simulation of the image overlaid with atoms illustrations. (f) STM topographic image across multiple steps of single layers, whose line profile is shown on the right. (g) Stacking alignment of two layers across a step, with the black and blue dots denoting Sn atoms of the upper and lower layers, respectively. All topographic data were taken at T=4.6K, V=-50mV, I=50pA.

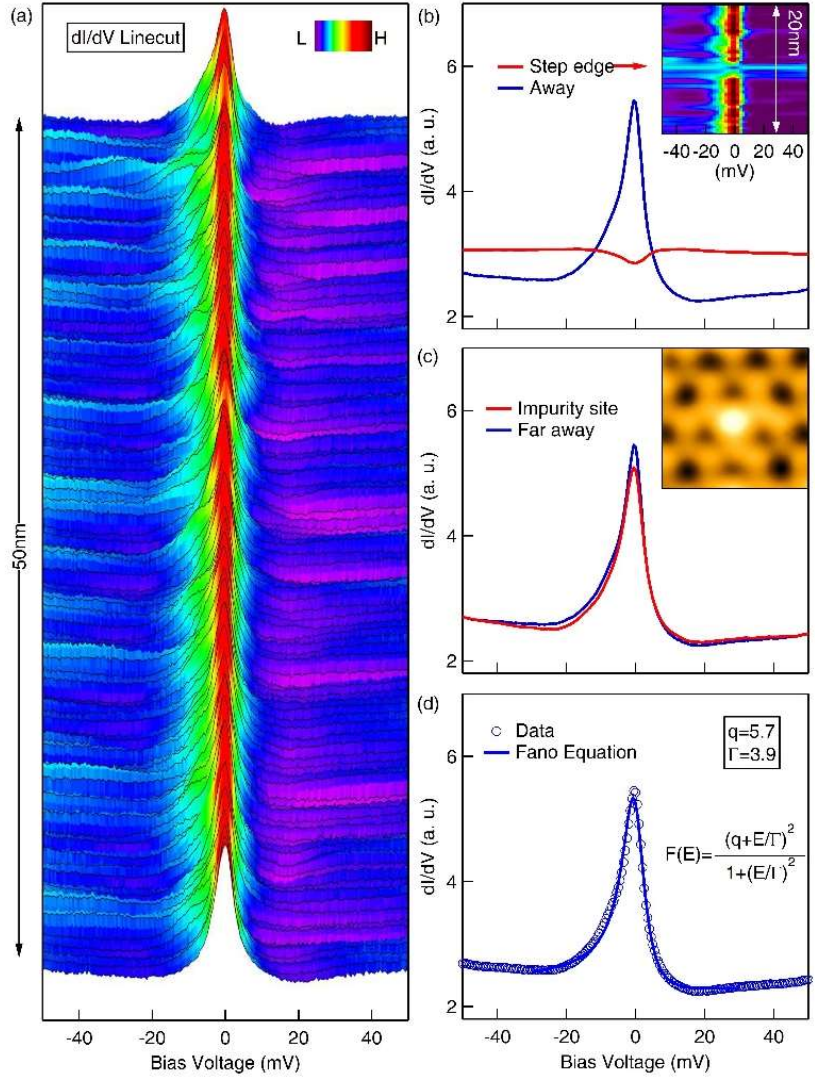

Fig. 2 (a) dI/dV line cut across 50nm, taken far from any step edge, showing a peak at $E_F$. Every fifth curve is marked by a solid black line for clarity. (b) Perturbation of a step edge on the Fermi level resonance. The inset shows the line-cut spectra taken across a single step edge, whose position is marked by the arrow. (c) Perturbation of the atomic impurity on the Fermi level resonance. The inset shows the topographic image of the impurity. (d) Spatially averaged dI/dV spectrum (open circles). The solid line is a fit to Fano line-shape function, $F(E) = \frac{(q+E/\Gamma)^2}{1+(\frac{E}{\Gamma})^2}$.

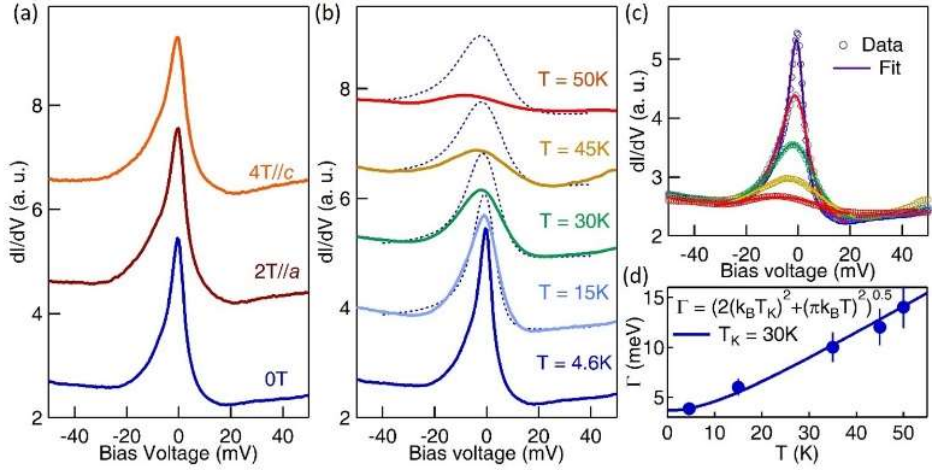

Fig. 3 (a) dI/dV taken at 0T (blue), 2T along a-axis (brown) and 4T along c-axis (yellow). (b) Temperature-dependence of dI/dV with spectra offset for clarity. The dotted lines are numerically calculated spectra by convoluting the 4.6K data with the derivative of Fermi-Dirac distribution function at respective temperatures. (c) Fit of the temperature-dependent spectra by thermally convoluted Fano function. (d) Resonance width $\Gamma$ plotted against temperature, giving a Kondo temperature of $T_K$=30K when fit to the Kondo model.

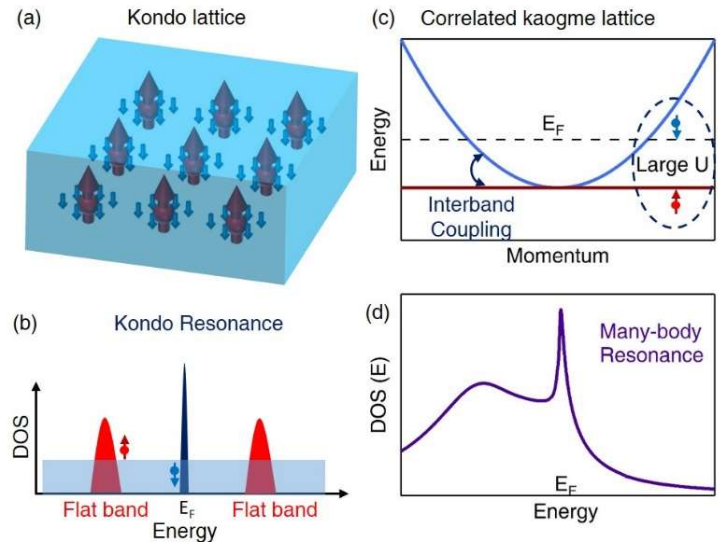

Fig. 4 (a) Schematic depicting a Kondo lattice formed by the coupling between periodic localized states (red arrows) and itinerant conduction electrons (blue arrows). (b) DOS spectrum of the Kondo lattice, where a Kondo resonance (dark blue) at $E_F$ is generated by the many-body coupling of the localized flat band (red) and itinerant conduction band (shaded blue). (c) Band structure for a kagome tight-binding model showing a flat band with interband coupling and a large Hubbard interaction. (d) The calculated DOS shows a weak bump arising from the kagome flat band and a many-body resonance at $E_F$ due to the hybridization of the localized and itinerant states.

**Acknowledgments**

Work at Princeton University was supported by the Gordon and Betty Moore Foundation (GBMF4547/ Hasan) and the United States Department of Energy (US DOE) under the Basic Energy Sciences programme (grant number DOE/BES DE-FG-02-05ER46200). S.S.T. and T.N. acknowledge support from the European Union's Horizon 2020 research and innovation programme (ERC-StG- Neupert-757867-PARATOP). T.-R.C. acknowledges MOST Young Scholar Fellowship (MOST Grant for the Columbus Program number 107-2636-M-006-004-), the National Cheng Kung University, Taiwan, and the National Center for Theoretical Sciences (NCTS), Taiwan. Z.W. is supported by the US Department of Energy, Basic Energy Sciences grant DE-FG02–99ER45747. This work is partially supported by CREST(JPMJCR18T3), Japan Science and Technology Agency, by Grants-in-Aids for Scientific Research on Innovative Areas (15H05882 and 15H05883) from the Ministry of Education, Culture, Sports, Science, and Technology of Japan, and by Grants-in-Aid for Scientific Research (19H00650) from the Japanese Society for the Promotion of Science (JSPS). M.Z.H. acknowledges support from Lawrence Berkeley National Laboratory and the Miller Institute of Basic Research in Science at the University of California, Berkeley in the form of a Visiting Miller Professorship.

**Supplementary Material**

**Materials and Methods:**

Single crystal Mn$_3$Sn samples were cleaved *ex situ* and then cleaned *in situ* with standard Ar ion sputtering and subsequently annealed at 1100K for approximately one hour before being inserted into the STM head at T = 4.6K after cooling to ~150K. Topographic images were obtained at T=4.6K, V=-50mV, and I=50pA. Differential conductance spectra were measured using a standard lock-in technique with modulation voltage V$_{RMS}$=0.3mV at V=-100~-50mV and I=500pA.

**First-principles calculation:**

The electronic structures were computed using the projector augmented wave method[28,29] as implemented in the VASP package[30] within the generalized gradient approximation scheme[31]. The spin−orbit coupling was included self-consistently in the calculations of electronic structures with a Monkhorst–Pack *k*-point mesh 21 × 21 × 21. Experimental lattice constants and the antiferromagnetic configuration were used[32]. The simulated STM image was based on the calculation of the real space charge density distribution according to the Tersoff-Hamann approach, which was acquired by the program HIVE[33]. In order to systematically calculate the surface and bulk electronic structure, we constructed a first-principles tight-binding model Hamiltonian for Mn$_3$Sn, where the tight-binding model matrix elements were calculated by projecting onto the Wannier orbitals, which used the VASP2WANNIER90 interface[34]. We used Mn *d*, and Sn *p* orbitals to construct Wannier functions without using the maximizing localization procedure. The surface states were calculated from the surface Green's function of the semi-infinite system. The band structure was renormalized by a factor of five to match the ARPES data[4].

**Resistivity measurements:**

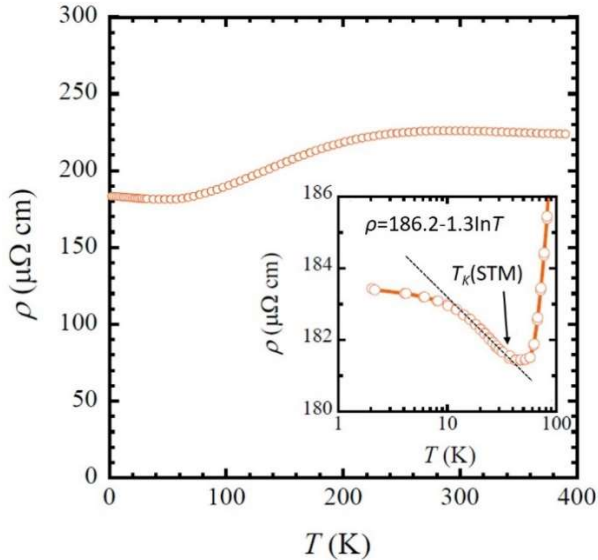

**Fig. S1** Resistivity measurements of bulk samples of Mn$_{3.07}$Sn$_{0.93}$ along the [2$\bar{1}\bar{1}$0] direction. The inset highlights the Kondo upturn.

**Kondo resonance in a kagome Hubbard model**

Since the electron correlation is indispensable in the system, we start with a 2D kagome Hubbard lattice model

$$H_f = -t \sum_{<i,j>,\sigma} d^\dagger_{i,\sigma} d_{j,\sigma} - \mu \sum_{i,\sigma} d^\dagger_{i,\sigma} d_{i,\sigma} + (U/2) \sum_{i,\sigma} n_{i,\sigma} n_{i,\bar{\sigma}} \quad (1)$$

where $t$, $\mu$ and $U$ are the hopping, the effective chemical potential, and the Hubbard U interaction respectively. $n_{i,\sigma} = d^\dagger_{i,\sigma} d_{i,\sigma}$. For U = 0, one can obtain the typical single-particle energy spectrum as shown in the left panel of Fig S2**a**. We treat the effective chemical potential as a tuning parameter. For $\mu = \mu_1$ as denoted by the orange dashed line in Fig.S2**a**, a single Fermi surface emerges together with a flat band below the Fermi pocket. For $\mu = \mu_2$, two Dirac bands need to be considered. We consider the first case $\mu = \mu_1$ while the second case can be found to generate qualitatively the same physical results with direct generalization.

Due to the presence of an onsite Hubbard interaction, the flat band degrees of freedom can also be essential although it is relatively far away from $\mu$. We project the lattice model to the low-energy window near the Fermi energy while keeping the flat band degrees of freedom. This results in a projected effective Hamiltonian, $H = H_c + H_f$, with

$$H_c = \sum_{\mathbf{k},\sigma} \varepsilon_{\mathbf{k}\sigma} c^\dagger_{\mathbf{k},\sigma} c_{\mathbf{k},\sigma} \quad (2)$$

and

$$H_f^0 = \epsilon_f \sum_{\mathbf{k},\sigma} f^\dagger_{\mathbf{k},\sigma} f_{\mathbf{k},\sigma} \quad (3)$$

where we use $c$ and $f$ to describe the itinerant band and flat band after projection respectively. A renormalization group analysis clearly shows that the projected Hubbard interaction $(U/2) \sum_{\mathbf{k},\mathbf{k}',\mathbf{q},\sigma} c^\dagger_{\mathbf{k},\sigma} c_{\mathbf{k}+\mathbf{q},\sigma} c^\dagger_{\mathbf{k}',\bar{\sigma}} c_{\mathbf{k}'-\mathbf{q},\bar{\sigma}}$ brings about only Landau Fermi liquid renormalization to the *c*-electrons with finite quasi-particle weight, therefore being irrelevant for the following analysis. Whereas the projected Hubbard interaction has crucial effects on the flat band whose kinetic energy is completely quenched due to the kagome lattice geometry, hence the total Hamiltonian describing the flat band is cast into

$$H_f = \epsilon_f \sum_{\mathbf{k},\sigma} f^\dagger_{\mathbf{k},\sigma} f_{\mathbf{k},\sigma} + (U/2) \sum_{\mathbf{k},\mathbf{k}',\mathbf{q},\sigma} f^\dagger_{\mathbf{k},\sigma} f_{\mathbf{k}+\mathbf{q},\sigma} f^\dagger_{\mathbf{k}',\bar{\sigma}} f_{\mathbf{k}'-\mathbf{q},\bar{\sigma}} \quad (4)$$

Furthermore, the interband scattering processes are unavoidable as manifestations of fluctuations of inter-sublattice scatterings. The most important scattering processes takes place between the electrons near the Fermi surface and those in the flat band, which is modeled by

$$H_{hyb} = V \sum_{\mathbf{k},\sigma} (c^\dagger_{\mathbf{k},\sigma} f_{\mathbf{k},\sigma} + h.c.). \quad (5)$$

To summarize, we expect that Eq. (2), (4), (5), which are reduced from the kagome Hubbard lattice model, include the major physics responsible for the experimental observation. The reduced model, resembling a periodic Anderson lattice model, has been widely studied and various of properties of ground state and phase diagrams have been discussed[35-44]. Here, we focus on (a) the Kondo resonance and discuss the results relevant to our experiment and (b) the effect of the spin-polarization of the *c*-fermion bath on the Kondo resonance which naturally exists in a non-collinear magnetic insulator.

The STM measurement probes the local density of states (LDOS) $\rho_f(\mathbf{r}, \omega) = -\left(\frac{1}{\pi}\right) \mathrm{Im} G_f^r(\mathbf{r}, \omega)$, where $G_f^r(\mathbf{r}, \omega)$ is the retarded Green's function (GF) of the *f*-fermions. To obtain $G_f^r(\mathbf{r}, \omega)$, we set up the equation of motion of the Green's functions up to the third order. In the following, we introduce the notation for the retarded Green's function as $\ll f_{r,\sigma}|f^{\dagger}_{r'\sigma}\gg \equiv -i\int_0^\infty dt <\{f_{r,\sigma}(t), f^{\dagger}_{r'\sigma}(t=0)\}> e^{izt}$, where $z = \omega + i\eta$ with $\eta$ being an infinitesimal positive number[45,46]. The retarded Green's function of two fermionic generic operators A,B satisfy the equation of motion $\omega \ll A|B \gg \equiv <\{A,B\}>_0 + \ll [A,H]|B \gg$, where $\{A,B\}$ is the anticommutation of the two operators. Starting from $\ll f_{r,\sigma}|f^{\dagger}_{r'\sigma} \gg$, one arrives at its equation of motion which generates new Green's functions. We continue setting up the equation of motion of the newly generated Green's functions, and after projecting, at the third-order expansion, the generated Green's functions onto $\ll f_{r,\sigma}|f^{\dagger}_{r'\sigma} \gg$, we make the whole coupled equations closed. Then, solving the coupled equations, one can obtain in the large U limit the **k**-space Green's function of f-electrons $G^r_f(\mathbf{k},\omega) = \frac{1}{N}\sum_{r,r'} \ll f_{r,\sigma}|f^{\dagger}_{r\sigma} \gg e^{i\mathbf{k}\cdot(\mathbf{r}-\mathbf{r'})}$, from which we calculated the LDOS of the f-electrons.

For a generic $\mathbf{r}$, the LDOS is shown by Fig.S2**b**. A robust peak is found for low temperatures, which is located very close to the Fermi energy (we considered both spin-degenerate and partially spin-polarized cases). The bump of the LDOS near $\varepsilon_f$ is a result of the large DOS from the exact flat spectrum of f-fermions, whose absence in real materials is expected, since the flat spectrum is replaced by quasi-flat bands in layered kagome lattices.

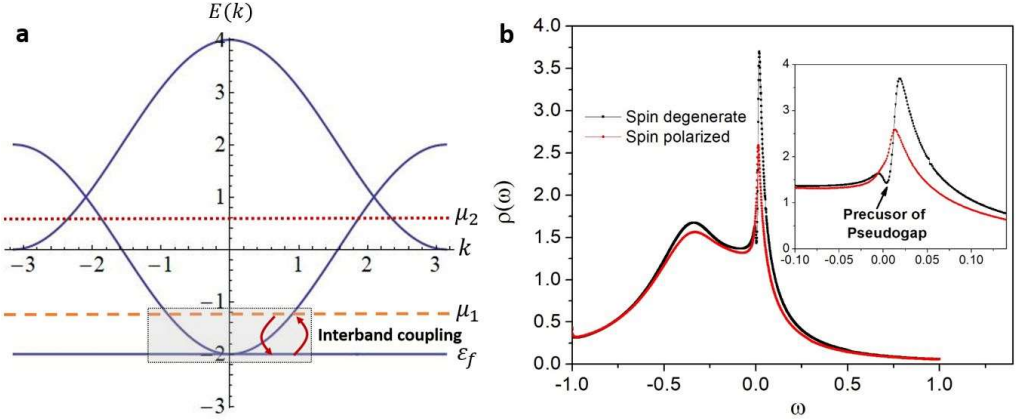

**Fig. S2 a,** The energy spectrum from the kagome lattice model with U = 0 and t < 0. **b** The LDOS of f-electrons for the large U limit for the spin degenerate and polarized case. V = 0.3, $\varepsilon_f = -0.3$, $\rho_0 = 0.5$ for the spin degenerate case while $\rho_{0\uparrow} = 0.5$ and $\rho_{0\downarrow} = 0.45$ for the polarized case. The inset shows the zoomed-in LDOS near Fermi energy, where a precursor of the pseudogap is observed for the spin degenerate case, which is suppressed after turning on the effective Zeeman field.